\documentclass[twocolumn]{aastex61}

\usepackage{graphicx}	
\usepackage{amsmath}	
\usepackage{amssymb}	

\newcommand{\Msun}{\,{\rm M_\odot}}

\newcommand\mybar{\kern1pt\rule[-\dp\strutbox]{.8pt}{\baselineskip}\kern1pt}

\shorttitle{Testing the Claimed Extremely Magnified Quasar at $z = 6.3$}
\shortauthors{Pacucci \& Loeb}

\begin{document}

\title{Reality or Mirage? Observational Test and Implications for \\ the Claimed Extremely Magnified Quasar at $z = 6.3$}

\correspondingauthor{Fabio Pacucci}
\email{fabio.pacucci@cfa.harvard.edu}

\author[0000-0001-9879-7780]{Fabio Pacucci}
\affil{Black Hole Initiative, Harvard University,
Cambridge, MA 02138, USA}
\affil{Center for Astrophysics $\vert$ Harvard \& Smithsonian,
Cambridge, MA 02138, USA}

\author[0000-0003-4330-287X]{Abraham Loeb}
\affil{Black Hole Initiative, Harvard University,
Cambridge, MA 02138, USA}
\affil{Center for Astrophysics $\vert$ Harvard \& Smithsonian,
Cambridge, MA 02138, USA}

\begin{abstract}
In the last two decades, approximately $200$ quasars have been discovered at $z>6$, hosting active super-massive black holes with masses $M_{\bullet} \gtrsim 10^9 \Msun$. While these sources reflect only the tip of the iceberg of the black hole mass distribution, their detection challenges standard growth models. The most massive $z>6$ black hole that was inferred thus far (J0100+2802, $M_{\bullet} \approx 1.2\times 10^{10} \Msun$) was recently claimed to be lensed, with a magnification factor $\mu=450$. 
Here, we perform a consistency check of this claim, finding that the detection of such a source requires a bright-end slope $\beta \geq 3.7$ for the intrinsic quasar luminosity function (LF), $\Phi(L) \propto L^{-\beta}$. Commonly used values of $\beta \sim 2.8$ are rejected at $>3\sigma$. 
If the claim is confirmed, it is very unlikely that all the remaining $51$ sources in the Sloan Digital Sky Survey sample are not magnified. Furthermore, it suffices that $\gtrsim 25\%$ of the remaining sources are lensed for the intrinsic LF to differ significantly (i.e., $>3\sigma$) from the observed one. The presence of additional extremely magnified sources in the sample would lower the requirement to $\sim 4\%$. Our results urge the community to perform more extended multi-wavelength searches targeting $z>6$ lensed quasars, also among known samples. This effort could vitally contribute to solving the open problem of the growth of the brightest $z\sim 7$ quasars.

\end{abstract}


\section{Introduction} \label{sec:intro}
The last two decades brought mounting evidence for the existence of super-massive black holes (SMBHs) in the very early Universe (e.g. \citealt{Fan_2003, Mortlock_2011, Wu_2015, Banados_2018}). The latest accounts \citep{Fan_2019} report the discovery of $\sim 200$ quasars (i.e., galaxies hosting an active SMBH), which are massive ($\gtrsim 10^9 \Msun$), very rare ($\sim 1 \, \mathrm{Gpc^{-3}}$) and formed very early in the cosmic history ($z \gtrsim 6$).
Current surveys are able to detect only very bright quasars, the tip of the iceberg of a significantly more numerous population of high-$z$ black holes, which encompasses a much wider black hole mass range ($\sim 10^{1-10} \Msun$).

The study of the earliest population of quasars is fundamental to our understanding of the high-redshift Universe. In fact, SMBHs significantly contributed to the formation and evolution of galaxies (e.g., \citealt{Kormendy_Ho_2013, Nguyen_2019, Shankar_2019}) and possibly played a role in the process of reionization (e.g., \citealt{Madau_2017}). Despite their importance, knowledge of early populations of SMBHs is limited. Most significantly, we do not know the initial conditions and the early evolution of the population, or how the first black holes formed and rapidly evolved into SMBHs by redshift $z\sim 7$ (e.g., \citealt{Woods_2019}). Questions concerning the rapid growth of SMBHs were already pointed out by \cite{Turner_1991} with the discovery of the first quasars at $4<z<5$ and became more pressing with the detection of $z > 6$ sources \citep{Haiman_Loeb_2001}. Currently, the farthest quasar is detected only $\sim 700 \, \mathrm{Myr}$ after the Big Bang ($z \approx 7.54$, \citealt{Banados_2018}).

Among the $z>6$ quasars, J0100+2802 stands out as the one hosting the most massive SMBH, $M_{\bullet} \approx 1.2\times 10^{10} \Msun$ at $z \approx 6.3$ \citep{Wu_2015}.
By stretching the parameter space of mass and time to an extreme, this quasar offers a unique view of the population of early SMBHs.
Assuming a continuous accretion with Eddington ratio $\lambda_{\mathrm{Edd}} \equiv \dot{M}/\dot{M}_{\mathrm{Edd}}$, the ratio between accretion rate and Eddington rate, and radiative efficiency $\epsilon_{\mathrm{rad}} \approx 0.1$, the growth time from the initial mass $M_{\mathrm{seed}}$ is
\begin{equation}
\tau_{\mathrm{growth}} \approx 0.45   \frac{\epsilon_{\mathrm{rad}}}{(1-\epsilon_{\mathrm{rad}})\lambda_{\mathrm{Edd}}} \ln{\frac{M_{\bullet}}{M_{\mathrm{seed}}} \, \mathrm{Gyr} \, .}
\end{equation}
Assuming that seeding occurs at $z \sim 30$ (e.g., \citealt{BL01}), the existence of this source requires \textit{continuous} accretion at the Eddington limit from a seed with a \textit{minimum} mass $M_{\mathrm{seed}} \sim 2,000 \, \Msun$. Albeit not theoretically forbidden, these extreme requirements are challenging. Several solutions were proposed to reduce the growth time, by either allowing for super-Eddington rates (e.g., \citealt{Begelman_1978, Wyithe_Loeb_2012}) and/or by increasing the initial mass of the seed (e.g., \citealt{Bromm_Loeb_2003}).

What if, instead, we are witnessing a ``mirage'', an optical illusion? The luminosity (and, consequently, the mass) of $z > 6$ quasars might be significantly over-estimated due to gravitational lensing by $z \lesssim 3$ galaxies. Depending on the slope of the intrinsic quasar luminosity function (LF), the lensing probability could be significant and close to unity \citep{Turner_1980, Comerford_2002, Wyithe_Loeb_2002}. Recently, \cite{Fan_2019} reported the discovery of a strongly lensed quasar at $z \approx 6.51$, with a magnification factor $\mu \sim 50$: the first detection of a lensed quasar at $z > 6$. Employing a lensing probability model calibrated on this detection, \cite{Pacucci_Loeb_2019} claimed that the observed population of reionization-era quasars contains several lensed sources with image separations below the resolution threshold of the \textit{Hubble Space Telescope} (HST). In addition, \cite{Fan_2019} pointed out that the quasar selection criteria currently employed are potentially missing a significant population of lensed quasars at $z > 6$, due to the fact that the lens galaxy contaminates the drop-out photometric bands. Assuming standard values of the slope of the quasar LF, \cite{Pacucci_Loeb_2019} claimed that the undetected quasars could account for up to $\sim 50\%$ of the known population.

Recently, \cite{Fugimoto_2019} studied J0100+2802 using data from the \textit{Atacama Large Millimeter Array} (ALMA), and found four separate and statistically significant peaks in the dust continuum map. They also report the detection of a Lya emission at $z \approx 2.3$, possibly identifiable as the lens galaxy. This could be indicative of a gravitationally lensed source, or of an ongoing merger event. If the source is lensed, the magnification factor in the optical would be $\mu \sim 450$, bringing the mass of the quasar below $\sim 10^9 \Msun$, more easily achievable with standard growth models. Previous analysis of the same ALMA data \citep{Wang_2019}, employing a different weighting method, reported no evidence of multiple peaks in the dust continuum map. 
These high-resolution observations are a cornerstone for studying the population of the earliest SMBHs in the Universe.

In this study, we test the hypothesis that J0100+2802 is lensed by $\mu \sim 450$ against current $z>6$ quasar data, constraining the value of the intrinsic quasar LF. Moreover, assuming that this source is magnified, we discuss the probability that additional quasars in the same sample are lensed. Finally, we quantify how the intrinsic quasar LF would differ from the observed one if a large number of $z>6$ quasars is lensed. Our calculations use the latest values of the cosmological parameters \citep{Planck_2018}.

\section{Lensing Model} 
\label{sec:theory}
We start with a summary of the lensing model employed in this study. See \cite{Pacucci_Loeb_2019} for a full description of the details.

\subsection{Lensing Probability}
To compute the cumulative magnification probability distribution, $P(>\mu)$, due to cosmologically distributed galaxies, we use the formalism presented in \cite{Pei_1995}.
The total magnification of a source at redshift $z_s$, produced by lenses at redshift $z'$, is labeled with $\mu$.
The probability distribution function for $\mu$ is
\begin{equation}
P(\mu) = \mu^{-1} \int_{-\infty}^{+\infty} \mathrm{d}s \exp{[-2\pi i s \ln{\mu} +Z(2 \pi i s|z_s)]} \, .
\label{eq:pdf}
\end{equation}
The moment function $Z(s|z_s)$ is defined as
\begin{equation}
\begin{split}
Z(s|z_s) = & \int_{0}^{z_s} \mathrm{d}z' \int_{1}^{\infty} \mathrm{d}A \rho(A, z'|z_s) \times \\
& \times (A^s -1 -0.4\ln{(10)}A +s) \, ,
\label{eq:moment}
\end{split}
\end{equation}
where the variable $A$ is related to the magnification $\mu$ via $\mu \equiv A/\bar{A}$, with $\bar{A}$ being the mean value of $A$.
For a source at redshift $z_s$, the quantity $\rho(A, z'|z_s)$ describes the mean number of lenses in the redshift range ($z'$, $z' + dz'$) and in the magnification range ($A$, $A + dA$).

Our model for the population of lensed sources assumes a double power-law shape for the quasar LF at $z > 6$. The faint-end slope and the break magnitude are fixed at $\alpha = 1.23^{+0.34}_{-0.44}$ and $M^*_{\rm 1450} = -24.90^{+0.75}_{-0.90}$, respectively \citep{Matsuoka_2018}, while the bright-end slope $\beta$ is variable: $\Phi(L) \propto L^{-\beta}$. The reason for this choice is that gravitational lensing is more significant for brighter sources, so the quasar LF is preferentially modified at its bright end.

\subsection{The Population and Distribution of Lens Galaxies}
The population of lenses is constituted of galaxies modeled as a truncated singular isothermal sphere and with flat rotation curves.
A full description of the statistical properties of this population is provided in \cite{Pacucci_Loeb_2019}. The physical properties of the lens galaxies are fully characterized by the dimensionless parameter $F = 3 \Omega_G/[2 r^2 D(z_s, z')]$ \citep{Pei_1993, Pei_1995}. Here, $\Omega_G$ is the cosmological density parameter of galaxies, and $D(z_s, z')$ is the angular diameter distance between source at $z_s$ and lens at $z'$. In addition, $r$ is the size parameter with dimensions of $\mathrm{length}^{-1/2}$, which is a function of the velocity dispersion of galaxies and expresses the physical extension of the lens (and, hence, its Einstein radius). We fix $F \sim 0.05$ \citep{Pei_1995} and verify that our results are unchanged within the full domain of interest $ 0.01<F < 0.1$.

The model needs to be supplemented with the cosmological distribution of lens galaxies, which are assumed to be distributed uniformly in space and of type E/S0.
We employ the Schechter function \citep{Schechter_1976} to model the UV LF for galaxies. This function correctly reproduces the distribution of galaxies for $z \lesssim 6$ (e.g., \citealt{Coe_2015}):
\begin{equation}
\Phi(L) = \frac{\Phi_{\star}}{L_{\star}} \left( \frac{L}{L_{\star}} \right)^{\alpha_g} \exp{\left(-\frac{L}{L_{\star}} \right)} \, ,
\end{equation}
where $\Phi_{\star}$ is the number density of galaxies of luminosity $L_{\star}$ (the break luminosity), and $\alpha_g$ is the faint-end slope.
We employ previous results on the UV LF for galaxies: \cite{Beifiori_2014} for $z < 1$, and \cite{Bernardi_2010} and \cite{Mason_2015} for $z \gtrsim 1$. As previously pointed out (e.g., \citealt{Wyithe_2011}), most of the lensing optical depth for $z \gtrsim 6$ sources is generated by lens galaxies at $z \lesssim 1.5$.
We use the Faber-Jackson relation ($L \propto \sigma_v^4$, \citealt{Faber_Jackson_1976}) to model the dependence of the velocity dispersion $\sigma_v$ on the luminosity of the galaxy.
The redshift evolution of the velocity dispersion is parametrized as $\sigma_v (z) \propto (1+z)^{\gamma}$, with $\gamma = 0.18 \pm 0.06$ \citep{Beifiori_2014}, suggesting a mild evolution (see also e.g. \citealt{Mason_2015}).

\section{RESULTS} 
\label{sec:results}
We are now in a position to test the claim that J0100+2802 is magnified by $\mu = 450$ against the known population of $z \gtrsim 6$ quasars. Furthermore, assuming this claim to be true, we present its theoretical implications on the broader population of high-$z$ quasars.

\subsection{Observational Test}
\label{subsec:test}
The $z=6.3$ quasar J0100+2802 was originally selected in the \textit{Sloan Digital Sky Survey} (SDSS) and reported by \cite{Wu_2015} as ultra-luminous. Applying a virial estimator based on the MgII line, its black hole mass was calculated as $\sim 1.2 \times 10^{10} \Msun$, making it the most massive $z > 6$ SMBH ever detected.

In the following, we assume that J0100+2802 is magnified by $\mu=450$ and calculate how many more lensed quasars we expect in the same SDSS survey. We require our prediction to be lower than the value inferred from the SDSS LF in each luminosity bin, as J0100+2802 is the only source that is claimed to be lensed.

\cite{Jiang_2016} used $N=52$ $z > 5.7$ quasars (including J0100+2802) to produce a $z \gtrsim 6$ quasar LF, deriving a slope $\beta = 2.8$.
We employ the same data to reconstruct their LF. We derive absolute magnitudes at $\lambda = 1450 \, \mathrm{\AA}$ from \cite{Jiang_2016} and employ the relevant bolometric correction from \cite{Runnoe_2012}. The number densities of the quasars are calculated using the standard $1/V_{\mathrm{max}}$ weighting method, where $V_{\mathrm{max}}$ is the maximum volume in which a given source is observable, assuming a survey nominal limit of $z_{\rm
AB} = 20.0$ mag for an uncertainty of $\sim 0.10$ mag in the SDSS. We divide the absolute magnitude range $-24 <M_{1450} < -29$ in seven bins to be consistent with \cite{Jiang_2016}.
\begin{figure}
\includegraphics[angle=0,width=0.5\textwidth]{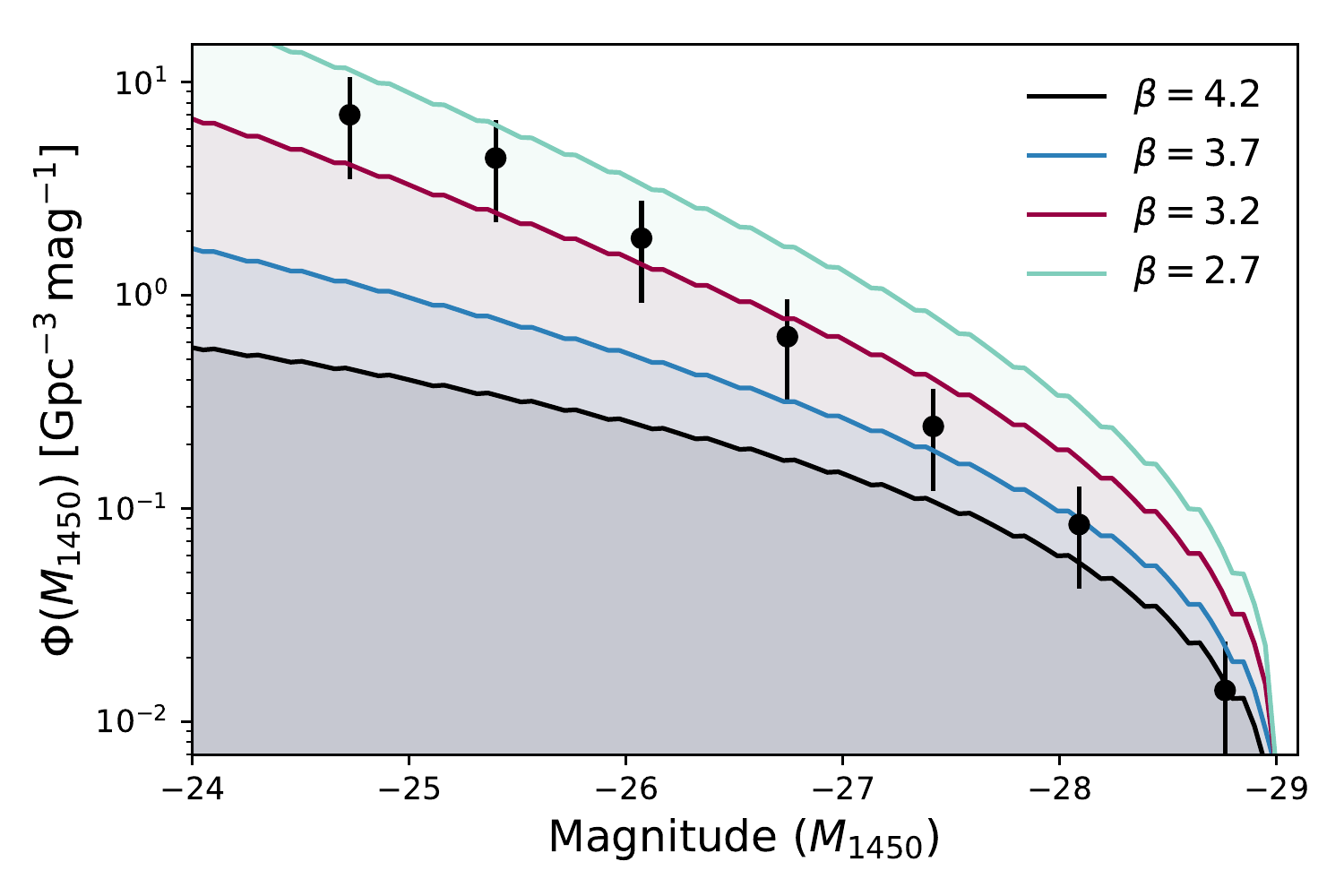}
\caption{The $z \gtrsim 6$ SDSS quasar LF calculated from \cite{Jiang_2016} is shown with black symbols and $1\sigma$ error bars. The lines show the expected number of lensed quasars assuming a bright-end slope of the intrinsic quasar LF of $\beta=2.7, 3.2, 3.7, 4.2$ as indicated in the legend. The lensing probability model \citep{Pacucci_Loeb_2019} assumes the detection of one quasar with $\mu = 450$ in the $M_{1450} \approx -29$ luminosity bin.} 
\label{fig:test}
\end{figure}

We then employ the lensing probability model in \cite{Pacucci_Loeb_2019} to calculate how many lensed quasars with $\mu < 450$ we expect in the same sample. The underlying assumption is that if J0100+2802 is intrinsically a black hole with $M_{\bullet} \sim 8\times 10^8 \Msun$ but magnified to appear as $M_{\bullet} \sim 10^{10} \Msun$ \citep{Fugimoto_2019}, then more of the fainter sources should be magnified by $\mu < 450$. We assume a flat distribution in intrinsic mass for the unlensed sources, as the observed luminosity depends only on the product between the magnification factor and the unlensed luminosity, which is proportional to the intrinsic mass. We use four values of the intrinsic quasar LF: $\beta=2.7, 3.2, 3.7, 4.2$. This range encompasses most of the values usually employed in literature (e.g., \citealt{Jiang_2016,Yang_2016}).

The results are shown in Fig. \ref{fig:test}. The models with $\beta=2.7$ and $\beta=3.2$ significantly overshoot the observed number density of $z \gtrsim 6$ quasars in several luminosity bins. Otherwise stated, if J0100+2802 is magnified by a factor $\mu=450$, we should have observed significantly more quasars at $M_{1450} \approx -27, -28, -29$. The prediction is inconsistent with SDSS data at $>3\sigma$. If the intrinsic quasar LF is instead steeper, with $\beta \geq 3.7$, the lensing probability model can still accommodate the presence of a $\mu=450$ quasar without overproducing the SDSS number counts. In summary, \textit{this consistency test strongly favours steep values ($\beta \geq 3.7$) for the intrinsic slope of the $z \gtrsim 6$ quasar LF}. Shallower values are significantly inconsistent with the presence of a source magnified by $\mu = 450$.

\subsection{Probability of Additional Lensed Sources \\ in the SDSS sample}
\label{subsec:more}
After checking that the presence of a quasar with $\mu = 450$ is, in principle, possible with steep values of the intrinsic quasar LF, we expand on the implications that its detection draws on the other sources in the sample. While we cannot a priori state which sources in the sample are lensed, we can derive strong predictions on the overall expected number of lensed sources and on their magnifications.

Following our results in Sec. \ref{subsec:test}, we fix $\beta = 3.7$ and compute the probability $P(<\mu) = 1 - P(\geq\mu)$ of having magnification factors $\mu_i < \mu$ for all the remaining $i = 1, \dotsc ,51$ sources in the sample. We denote this ensemble probability, $P_{\mathrm{SDSS}}$. We also assume that each of the remaining quasars are represented by stochastic, independent variables: once the overall lensing model is fixed, the magnification factor of one source in the sample is independent from all the others. The probability $P_{\mathrm{SDSS}}$ is then
\begin{equation}
P_{\mathrm{SDSS}} = \prod_{i=1}^{51} [1 - P_i(\geq\mu)] \, .
\end{equation}

\begin{figure}
\includegraphics[angle=0,width=0.5\textwidth]{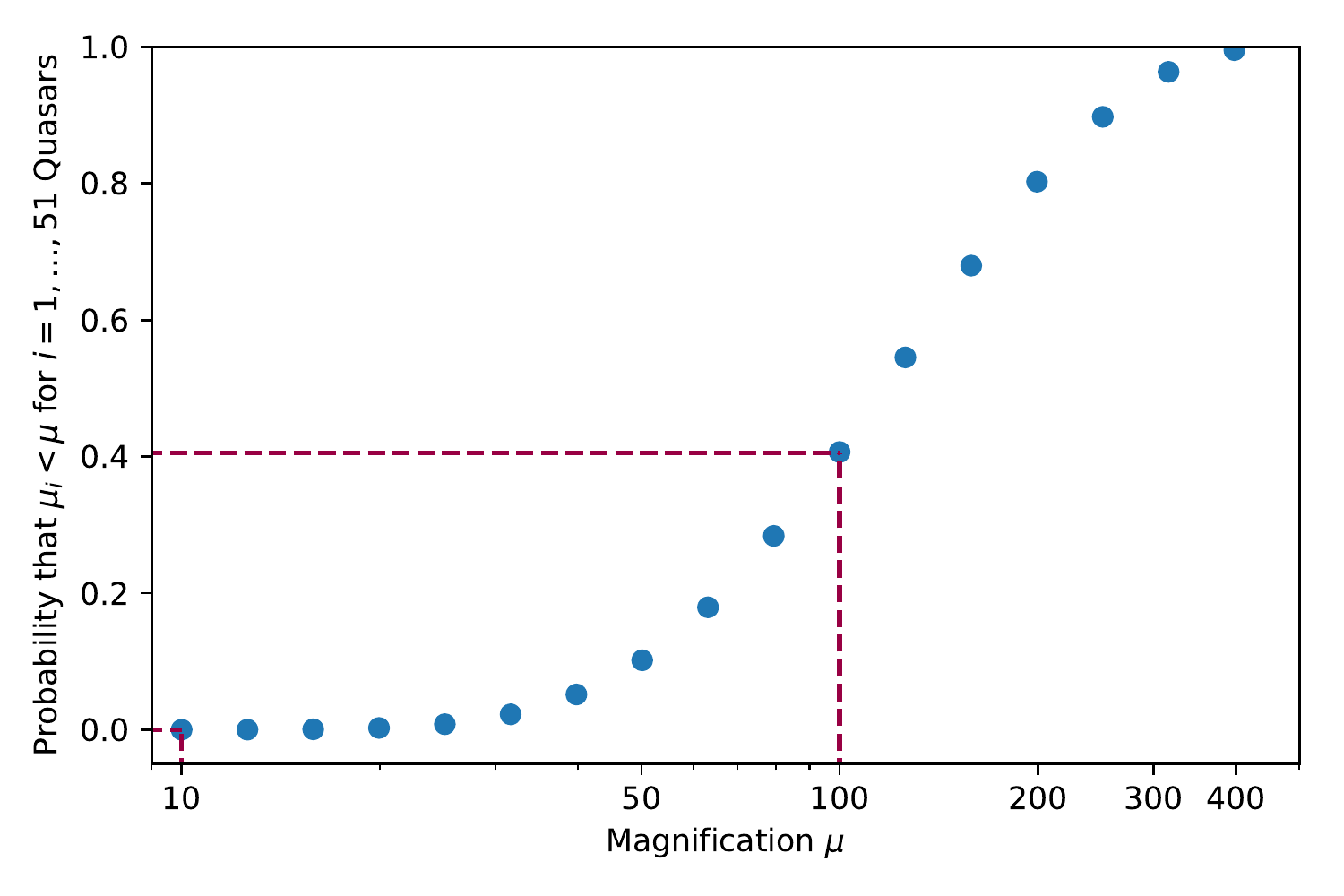}
\caption{Probability $P_{\mathrm{SDSS}}$ that all the remaining $51$ sources in the SDSS sample have magnifications $\mu_i<\mu$. Assuming the presence of one source with $\mu = 450$, it is nearly impossible, e.g. $P(\mu < 10) \sim 10^{-5}$, that all of the remaining sources are not magnified. The values discussed in the text are indicated with red dashed lines.}
\label{fig:low_prob}
\end{figure}

The result is shown in Fig. \ref{fig:low_prob}. The probability that all the remaining $51$ sources have $\mu_i < 10$ is $P(\mu_i < 10) \sim 10^{-5}$. If J0100+2802 is magnified by $\mu = 450$ there is almost certainty that there is at least one quasar magnified with $\mu \geq 10$ in the same sample. Even more interestingly, Fig. \ref{fig:low_prob} conveys that $P(\mu_i < 100) \sim 0.4$, i.e., there is a $\sim 60\%$ chance that at least another quasar in the sample is extremely magnified, with $\mu \geq 100$.

If the claim about J0100+2802 is confirmed, these probability calculations will serve two purposes in guiding future searches for lensed $z>6$ quasars. From a theoretical perspective, they indicate that the extremely large mass of some high-$z$ quasars could be an optical illusion, thus leading to a re-consideration of current black hole growth models. From an observational point of view, the high probability that additional SDSS quasars are strongly lensed motivates future searches for these sources, with higher resolution observations either employing longer wavelengths (e.g., ALMA) or next-generation telescopes, such as the \textit{Wide Field Infrared Survey Telescope} (WFIRST) and the \textit{James Webb Space Telescope} (JWST).

In particular, we forecast that WFIRST will be instrumental for discovering a large number of $z \gtrsim 6$ quasars, owing to its deep wide-area survey capabilities. Following the mission specifications (see \citealt{Spergel_2015} and the \href{https://wfirst.gsfc.nasa.gov/}{WFIRST website}), it is instructive to estimate the number of lensed $z\gtrsim 6$ quasars that this survey could detect. For the planned High Latitude Survey, we assume a survey area $A_{\mathrm{WFIRST}} \sim 2,200 \, \mathrm{deg^2}$ reached to a limiting magnitude $m_{\mathrm{lim,AB}} \sim 27$. We then integrate the LF over luminosity and redshift to find the expected number $N_{\mathrm{WFIRST}}$ of $z > z_0$ quasars discoverable by WFIRST:
\begin{equation}
N_{\mathrm{WFIRST}} (z>z_0) = \int _ {L_{\rm lim}} ^ {+\infty} d \mathrm{L} \int _ {z_{\rm 0}} ^ {+\infty} \frac{d \mathrm{V}}{d \mathrm{z}} d \mathrm{z} \, \Phi ( L , z ) \, ,
\end{equation}
where ${L_{\rm lim}}$ is the bolometric luminosity of a quasar at redshift
z corresponding to an apparent magnitude $m_{\rm lim, AB}$ and $\mathrm{V}$ is the comoving volume. We obtain $N_{\mathrm{WFIRST}}(z > 6) \sim 5\times 10^4$, while restricting our attention to $z > 7$ we obtain $N_{\mathrm{WFIRST}}(z > 7) \sim 3,000$.
For this order-of-magnitude estimate, we employ the \cite{Jiang_2016} LF for sources with $M_{\rm 1450} < -24$ and the \cite{Matsuoka_2018} LF for sources with $M_{\rm 1450} > -24$. 
The evolution of the LF for $z\gtrsim 6.5$ is uncertain: for the purpose of this estimate, we keep the shape of the $z\gtrsim 6.5$ LF equal to the one presented by \cite{Jiang_2016}, while we change the overall density of quasars following the prescription by \cite{Wang_2019_density}. Note that the spatial density of quasars is expected to decline by $\sim 1$ order of magnitude between $z=6.5$ and $z = 8$.
Assuming a conservative value of $\beta = 2.8$, our lensing model predicts that the WFIRST $z > 6$ quasar sample could contain $\sim 500$ quasars with magnification factors $\mu \geq 10$ and $\sim 50$ quasars with magnification factors $\mu \geq 100$. Irrespective of the reality of the claim about J0100+2802, WFIRST will certainly play a crucial role in investigating the putative population of lensed high-$z$ quasars.

\subsection{Toward the Intrinsic Quasar LF}
Building upon our results in Sec. \ref{subsec:more}, we now aim to investigate how the presence of an additional number $i < N = 52$ of lensed quasars in the sample modifies the observed $z > 6$ quasar LF.

Previous studies (e.g., \citealt{Turner_1980, Wyithe_Loeb_2002, Wyithe_2002b, Wyithe_2011}) already pointed out that the intrinsic LF for quasars could significantly differ from the observed one. In fact, magnification bias could artificially increase the number counts of bright objects, thus affecting the bright-end slope of the LF. For example, \cite{Wyithe_2002b} placed a limit on the slope of the $z \sim 6$ quasar LF $\beta \lesssim 3$ using the fact that none of the quasars found by \cite{Fan_2003} are strongly lensed (i.e., $\mu > 2$). Note that in Sec \ref{subsec:test} we placed a limit $\beta \geq 3.7$ for the intrinsic LF instead, due to the putative detection of one quasar with $\mu = 450$.

To investigate how the presence of an additional number $i < N = 52$ of lensed quasars in the sample modifies the observed LF, we proceed as follows. Assuming the lensing probability model from \cite{Pacucci_Loeb_2019} with $\beta = 3.7$, for each integer $i < 52$ we draw a number $i$ of magnification factors from the given probability distribution. We then decrease the observed luminosities $L_{1450}$ with the appropriate magnification factors for the $i$ randomly chosen lensed sources. We re-construct the LF accounting for all the $i$ sources which have changed luminosity. Finally, we compare the new, de-lensed, LF with the one observed by \cite{Jiang_2016} employing the statistics
\begin{equation}
\Sigma(i) = \sqrt{\sum_{k=1}^{n_{\mathrm{bins}}} \frac{(C_{k, \mathrm{obs}} - C_{k, \mathrm{mod}})^2}{\sigma_k^2} } \, ,
\label{eq:statistics}
\end{equation}
where $n_{\mathrm{bins}}$ is the number of luminosity bins used to construct the LF, $C_{k, \mathrm{obs}}$ and $C_{k, \mathrm{mod}}$ are the observed and the modified source counts in the $k$-th bin, respectively, and $\sigma_k$ is the error associated with each measure in the observed LF \citep{Jiang_2016}.
As the magnification factors are randomly drawn from a probability distribution, the occurrence of large values of $\mu$ has a very significant impact on the statistics in Eq. (\ref{eq:statistics}). To smooth out random fluctuations, we run the experiment described above $10^6$ times, obtaining the average effect. We checked that the average line is insensitive to a number of trials larger than $10^6$.

\begin{figure}
\includegraphics[angle=0,width=0.5\textwidth]{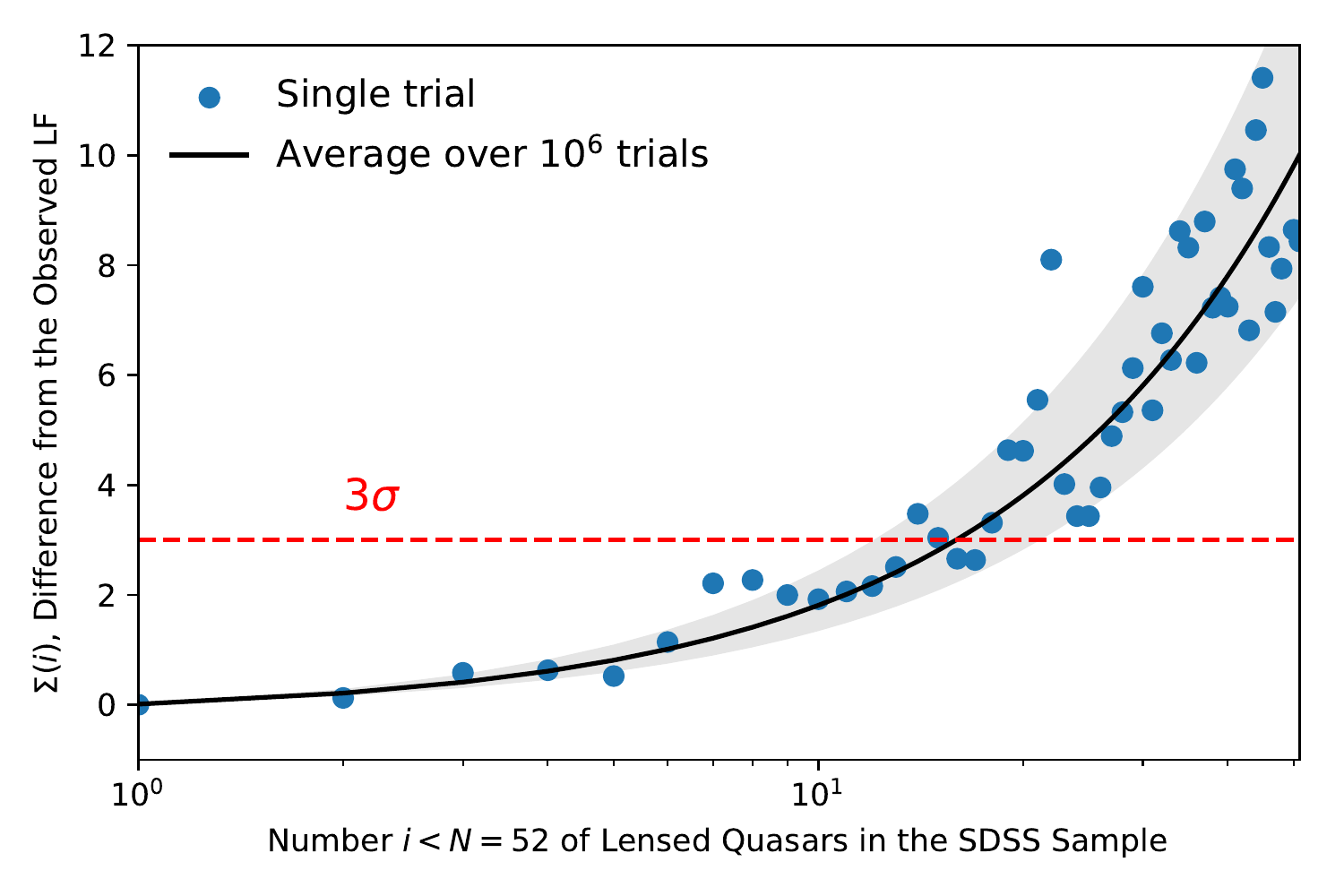}
\caption{Plot showing how the intrinsic LF would differ (in terms of the standard deviation $\sigma$) from the SDSS one \citep{Jiang_2016} if an additional number $i<52$ of quasars in the sample are lensed. Blue points are the results for a single trial, the black line is the average over $10^6$ trials, the shaded region indicates the $1\sigma$ uncertainty.}
\label{fig:change_LF}
\end{figure}

The result for $\Sigma(i)$ is shown in Fig. \ref{fig:change_LF}.  We predict that, on average, a number $i > 15 \pm 3$ of lensed quasars among the $N=52$ SDSS sample ($\gtrsim 25\%$ of the sample) modifies the bright-end slope of the quasar LF by $> 3\sigma$ with respect to the observed one. 
Of course, it is possible that even a small number $i \ll 15$ of strongly lensed sources can significantly affect the LF, especially if they all occur in the same luminosity bin. The presence of additional extremely magnified sources would require a sample fraction as low as $\sim 4\%$ to reach an inconsistency $> 3\sigma$ with respect to the observed LF.

\vspace{0.4cm}

\begin{figure}
\includegraphics[angle=0,width=0.5\textwidth]{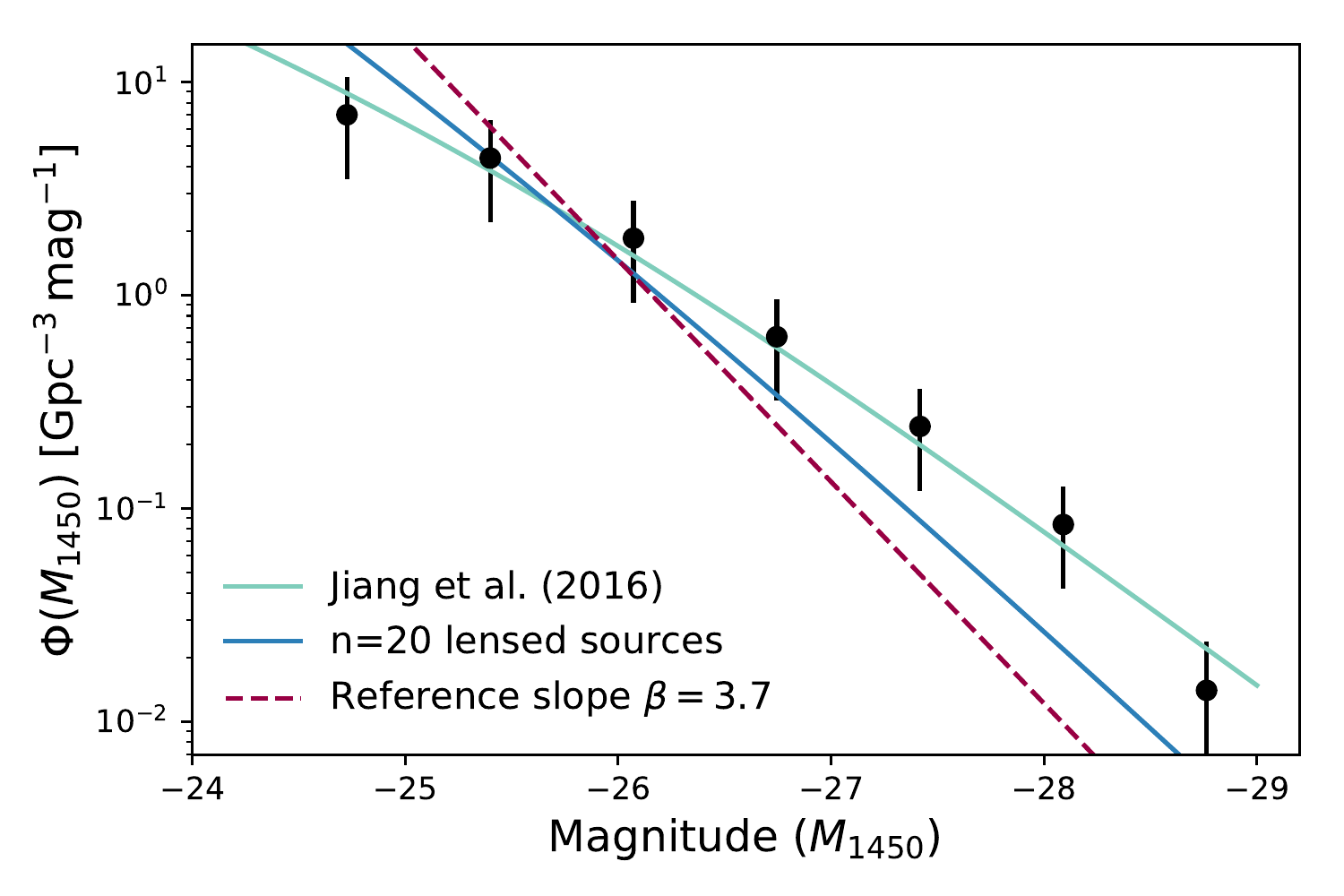}
\caption{Example of the lensing effect on the quasar LF for a number $i=20$ of lensed quasars in the SDSS sample. Black symbols are data from \cite{Jiang_2016}; the green line being their best fit, and the blue line being the quasar LF that we obtain assuming that additional $20$ randomly chosen quasars in the sample are magnified. The red dashed line indicates a slope $\beta=3.7$, for reference.}
\label{fig:Fig4}
\end{figure}

A practical example of how a number $i<52$ of lensed quasars in the SDSS sample would modify the observed quasar LF is shown in Fig. \ref{fig:Fig4}. The green line is the best fit to the SDSS sample from \cite{Jiang_2016}, shown with black symbols, using a double power-law fitting function with $\alpha = 1.9$ and $\beta = 2.8$. The blue line is a double power-law fit to a modified sample obtained assuming that $i=20$ randomly chosen quasars from the SDSS sample are magnified. The magnification factors are drawn from the lensing probability distribution described in \cite{Pacucci_Loeb_2019} with a value of $\beta = 3.7$ for the intrinsic quasar LF. This fitting line is significantly inconsistent with data from \cite{Jiang_2016} at the bright end. In fact, as more quasars are magnified, their true luminosity is decreased and their contribution to the quasar LF is shifted to the faint end. For reference, the red dashed line indicates a bright-end slope $\beta = 3.7$.

\vspace{0.4cm}

\section{Discussion and Conclusions} 
\label{sec:disc_concl}
Our study is motivated by the claim \citep{Fugimoto_2019} that J0100+2802, a $z \approx 6.3$ quasar with $M_{\bullet} \approx 1.2\times 10^{10} \Msun$ \citep{Wu_2015} is lensed by a magnification factor $\mu=450$.
The black hole mass is calculated via a virial estimator based on the MgII line which scales as 
$M_{\bullet} \propto (\lambda L_{\lambda, \mathrm{3000}})^{0.5}$ \citep{Fugimoto_2019}, such that the final, unlensed mass would be $\sim 8\times 10^8 \Msun$, decreasing the growth time by $\sim 20\%$. It is worth noting that also the Eddington ratio for J0100+2802 would change dramatically, from $\lambda_{\mathrm{Edd}} \approx 1$ to $\lambda_{\mathrm{Edd}} \approx 0.07$. 

This source could be the second $z > 6$ lensed quasar found in less than one year \citep{Fan_2019} and, by far, the one with the highest magnification.
A magnification factor $\mu = 450$ has a probability $P(>450) \sim 10^{-3}$ to occur in the most favorable lensing models (see \citealt{Pacucci_Loeb_2019} with $\beta = 3.6$): claiming it in a relatively small sample of sources invites skepticism, but we might have observed an extraordinary source such as J0100+2802 only because it is extremely magnified.
It is thus instructive to understand the theoretical implications of this putative detection.

We performed a consistency check of the \cite{Fugimoto_2019} claim, finding that a detection of one quasar with $\mu=450$ in the SDSS sample \citep{Jiang_2016} requires a slope $\beta \geq 3.7$ for the intrinsic quasar LF. Commonly used values of $\beta \sim 2.8$ are rejected at $>3\sigma$.

Assuming that the claim is real, we derive that it is nearly impossible that all the remaining $n=51$ sources in the SDSS sample are not magnified by at least $\mu = 10$. Furthermore, on average, it is sufficient that $\gtrsim 25\%$ of the remaining sources in the SDSS sample are lensed for the intrinsic LF to differ significantly (i.e., $>3\sigma$) from the observed one. The presence of additional extremely magnified sources in the sample would even require a much smaller percentage, as low as $\sim 4\%$.

It is worth noting that the consistency check performed in Sec. \ref{subsec:test} does not necessarily rule out intrinsic slopes $\beta < 3.7$. In fact, if the prediction on the number density of lensed quasars falls significantly above the value calculated from the SDSS LF, then we expect (at least) one of the following statements to be true: (i) J0100+2802 is not magnified by $\mu=450$; (ii) we are missing a significant fraction of high-$z$ lensed quasars in current LFs \citep{Pacucci_Loeb_2019}.
Regarding point (i), several HST snapshot surveys were already performed on high-$z$ quasars \citep{Maoz_1993, Richards_2004, McGreer_2013_snapshot}, resulting in no lensed quasar candidates.
Regarding point (ii), it is possible that many lensed objects in the SDSS sample have image separations smaller than the HST resolution ($\sim 0.''1$), leading to a missed identification of the lens. Previous studies (e.g., \citealt{Keeton_2005, Pacucci_Loeb_2019}) have estimated the probability of this event occurring for $z>6$ quasars at $\sim 20\%$.

It is thus of utmost importance to first test observationally the claim that J0100+2802 is magnified by $\mu=450$. 
Thus far, the strongest argument against the lensing hypothesis comes from the very extended proximity zone around J0100+2802.
The proximity zone is defined as the physical region around the quasar inside which its Ly$\alpha$ flux is above 10\% of its peak value (e.g., \citealt{Fan_Carilli_2006}). The behavior of the quasar proximity zone is thoroughly described in literature \citep{Shapiro_1987, Cen_Haiman_2000, Madau_Rees_2000, Haiman_2005, Fan_Carilli_2006, Carilli_2010, Eilers_2018} and is a powerful probe of the intrinsic ionizing power of a quasar.
In the discovery paper, \cite{Wu_2015} already mentioned that the quasar proximity zone for J0100+2802 is as large as $\sim (7.9 \pm 0.8) \, \mathrm{Mpc}$. Such an extended ionized region likely requires a very massive and active quasar to form. 
Note that the proximity zone was previously used by \cite{Haiman_Cen_2002} to rule out the possibility that a quasar at $z=6.28$ is lensed and, more recently, by \cite{Fan_2019} to confirm that a quasar at $z = 6.51$ is strongly lensed.

To verify the claim, deeper ALMA observations are needed, as the HST does not seem to discern multiple images of this source \citep{Fugimoto_2019}. We urge the community to check this claim and to look for additional lensed $z > 6$ quasars, possibly using a multi-wavelength approach, as current optical/infrared observations might not be able to discern multiple sources \citep{Pacucci_Loeb_2019}. Next-generation telescopes will likely play a primary role in this search. In particular, we showed that WFIRST has the potential of discovering $5\times 10^4$ new quasars at $z > 6$, and $\sim 3000$ at $z > 7$. Current lensing models suggest that many of them could be strongly lensed: $\sim 500$ with $\mu \geq 10$. If this is the case, the high-resolution power of WFIRST and JWST will certainly be instrumental for confirming their lensed nature, by identifying multiple images. The confirmation that J0100+2802 is lensed and/or the detection of more lensed sources could vitally contribute to solve the open problem of the growth of the first SMBHs, already detected by $z\sim 7$, less than $10^9$ yr after the Big Bang.

\vspace{0.3cm}
We thank the anonymous referee for constructive comments on the manuscript.
F.P. acknowledges fruitful discussions with X. Fan, F. Wang, and S. Fujimoto, as well as with several participants of the workshop ``Accretion History of AGN 2019'' in Miami, Florida, including N. Cappelluti, A. Comastri, R. Gilli, R. Hickox, E. Treister, and F. Vito.
This work was supported by the Black Hole Initiative at Harvard University, which is funded by grants from the John Templeton Foundation and the Gordon and Betty Moore Foundation.





\bibliographystyle{mnras}
\bibliography{ms}
\label{lastpage}
\end{document}